\documentclass[preprintnumbers,amsmath,amssymb,floatfix,nofootinbib]{revtex4}
\usepackage{graphicx}
\usepackage{dcolumn}
\usepackage{bm}
\usepackage{amsthm}
\usepackage{amsmath}
\usepackage{amsfonts}
\usepackage{color}

\usepackage{caption}

\newcommand{\be}{\begin{equation}}
\newcommand{\ee}{\end{equation}}
\newcommand{\ba}{\begin{eqnarray}}
\newcommand{\ea}{\end{eqnarray}}

\newcommand{\n}[1]{\label{#1}}
\newcommand{\non}{\nonumber}
\newcommand{\eq}[1]{(\ref{#1})}

\newcommand{\hu}{\widehat{U}}
\newcommand{\hw}{\widehat{W}}

\newcommand{\hv}{\widehat{V}}

\newcommand{\hh}{\, ,\hspace{0.5cm}}

\newcommand{\bi}[1]{\bibitem{#1}}

\newcommand{\BM}[1]{{\mbox{\boldmath $#1$}}}

\begin{document}

\title{\boldmath Thermodynamic of Distorted Reissner-Nordstr\"om Black Holes in Five-dimensions}

\author{Shohreh Abdolrahimi}
\email{shohreh.abdolrahimi@uni-oldenburg.de}
\affiliation{Institut f\"ur Physik, 
Universit\"at Oldenburg, 
Postfach 2503 D-26111 Oldenburg, Germany}

\begin{abstract}
In this paper, we study mechanics and thermodynamics of distorted, five-dimensional, electrically charged (non-extremal) black holes on the example of a static and ``axisymmetric'' black hole distorted by external, electrically neutral matter. Such a black hole is represented by the derived here solution of the Einstein-Maxwell equations which admits an $\mathbb{R}^1\times U(1)\times U(1)$ isometry group. We study the properties of this distorted black hole.
\end{abstract}
\maketitle

\section{Introduction}
This is based on a paper \cite{ShASh}, which has been presented as a talk in the Karl Schwarzschild Meeting in Frankfurt. Einstein equations are very complex and describing black hole interaction with external matter and fields  usually requires involved numerical computations. To construct exact solutions which would model to some extend the interaction of a black hole with the external matter, Geroch and Hartle \cite{Ger.IV} proposed to consider static (or stationary), axisymmetric space-times which are not asymptotically flat. Such solutions represent black holes distorted by external matter. In the case of static and vacuum, axisymmetric 4-dimensional space-times, the external metric near these distorted black holes is given by a Weyl solution \cite{Weyl.IV}. Such solution represents, what is called, a local black hole solution, which is a space-time metric in the external neighborhood of a distorted black hole horizon. The external matter is not included into the solution and ``located" at the asymptotic infinity. 

Four-dimensional, distorted, static, axisymmetric, vacuum black holes were studied in, e.g., \cite{IsKh.IV,MS,W1,Chandra,Fai.IV,FS1}. Distorted, static, axisymmetric, electrically charged black holes were studied in, \cite{W3,Fai.IV,AFS.IV}, and distorted, stationary, axisymmetric, vacuum and electrically charged black holes were studied in, e.g., \cite{Tomi,Bre.IV} and \cite{Bre2.IV,Hen1,Hen2,Hen3}, respectively. Distorted black holes can show some strange and remarkable properties.   

In four dimensions, the general static, axisymmetric solution of the vacuum Einstein equations can be written in the form of Weyl solution which is defined by one of its metric function solving the Laplace equation in an auxiliary three-dimensional flat space. The other remaining metric function is derived by a line integral defined by the first one. Having written one of the Einstein equations as the Laplace equation
in a three-dimensional flat space is a great advantage. A distorted black hole solutions can be constructed by adding extra harmonic functions to the original one, which represents the black hole source. These functions represent the distortion fields. As it was done in the case of four-dimensions, one can use the generalized Weyl solution \cite{Emp.IV} to construct distorted black objects by adding the distortion fields to the Newtonian potentials, which define the solution. Here, we shall construct and analyze a solution representing a distorted five-dimensional Reissner-Nordstr\"om black hole. This is a static solution of the Einstein-Maxwell equations which has $\mathbb{R}^1\times U(1)\times U(1)$ isometry group. The construction is based on the gauge transformation of the matrix which is an element of the coset target space $SL(2,\mathbb{R})/U(1)$ of the scalar fields which define our model (see, e.g., \cite{GX}). 

Having our solution constructed, we shall analyse properties of such a black hole, focusing on its outer and inner horizons and the interior region located between them. We compare the distorted black hole solution with the Reissiner-Nordstr\"om solution representing undistorted black hole and study which properties remain and which are changed due to the distortion. 
 
In this paper, we use the following convention of units: $G_{(5)}=c=\hbar=k_{B}=1$, where $G_{(5)}$ is the five-dimensional gravitational constant. The space-time signature is +3 and the sign conventions are that adopted in \cite{MTW}.

\subsection{The generating transformation}
The metric of a five-dimensional static spacetime can be written in the following form:
\be\n{3.1}
ds^2=-e^{2U}dt^2+e^{-U}h_{ij}dx^idx^j\,,
\ee
where $U=U(x^{i})$, $h_{ij}=h_{ij}(x^i)$,  $i,j=(1,2,3,4)$. Einstein-Maxwell equations can be derived from the following action: 
\be\n{3.4}
\mathcal{S}=\frac{1}{16\pi}\int d^4x \sqrt{h}\left(\mathcal{R}+\frac{3}{4} {h}^{ij}\text{Tr}\left\{A_{,i} (A^{-1})_{,j}\right\}\right)\,,
\ee 
where $h=det(h_{ij})$, $\mathcal{R}$ is the four-dimensional Ricci scalar defined by the metric $h_{ij}$, and  
\be\n{3.5}
A=\begin{pmatrix} e^{U}-\frac{\Phi^{2}}{{3}}e^{-U}&&&&-\frac{\Phi}{\sqrt{3}}e^{-U}\\
-\frac{\Phi}{\sqrt{3}}e^{-U}&&&&-e^{-U}\end{pmatrix}\in SL(2,\mathbb{R})/U(1)\,.
\ee
 The action \eq{3.4} is invariant under the symmetry transformation
 \be 
 \bar{A}=GA\,G^T\,,
\ee
where $G\in SL(2,\mathbb{R})$ is a constant matrix. 
To illustrate this transformation let us consider a static asymptotically flat vacuum space-time of the form \eq{3.1}, then
\be\n{3.12} 
A=\begin{pmatrix} e^{U}&&&0\\ 
0&&&-e^{-U}\end{pmatrix}\,.
\ee
Applying the transformation 
\be\n{3.11}
G=\begin{pmatrix} \cosh\delta &&&\sinh\delta\\
\sinh\delta &&&\cosh\delta\end{pmatrix}\in SO(1,1)\subset SL(2,\mathbb{R})\,,
\ee
we derive
\ba
ds^2&=&-e^{2\bar{U}}dt^2+e^{-\bar{U}}
h_{ij}dx^idx^j\,,\n{3.13}\\
e^{\bar{U}}&=&\frac{e^{U}}{(\cosh^2\delta-e^{2U}\sinh^2\delta)}\,,\n{3.14}\\
\bar{\Phi}&=&-\sqrt{3}
\frac{(e^{2U}-1)\tanh\delta}{(1-e^{2U}\tanh^2\delta)}\,,\n{3.15}
\ea
where $\Phi$ is the electrostatic potential. This is a static solution of the five-dimensional Einstein-Maxwell equations. We apply the generating  transformation presented above to a five-dimensional Schwarzschild-Tangherlini black hole distorted by static and neutral, distribution of external matter, which creates only gravitational field \cite{AS-S}. The sources of the distortion (the external matter) are located at the asymptotic infinity and not included into the solution. The distorted five-dimensional vacuum black hole solution is not asymptotically flat. Therefore, the sought distorted five-dimensional electrically charged black hole solution is not expected to be asymptotically flat either. 
Consider the distorted five-dimensional Schwarzschild-Tangherlini solution in the form \eq{3.1} suitable for the transformation,
\ba
&&\hspace{2cm}ds^2=-e^{2U}dt^2+e^{-U}h_{ij}dx^idx^j\hh e^U=\sqrt{\frac{\eta-1}{\eta+1}}e^{\hu+\hw}\,,\non\\ 
&&\hspace{3.5cm}h_{ij}dx^idx^j=\frac{m}{4}
\sqrt{\eta^2-1}e^{\hu+\hw}dl^2\,,\n{4.1}\\
&&\hspace{0cm}dl^2=e^{2(\hv+\hu+\hw)}
\left(\frac{d\eta^2}{\eta^2-1}+d\theta^2\right)
+2(1+\cos\theta)e^{-2\hw}d\chi^2
+2(1-\cos\theta)e^{-2\hu}d\phi^2\,.\non
\ea
Here $\hu$, $\hw$, and $\hv$ are the distortion fields given by the following expressions: 
\ba
\hu(\eta,\theta)&=&\sum_{n\geq0}a_n\,
R^{n}P_n\hh\hw(\eta,\theta)=\sum_{n\geq0}b_n\,
R^{n}P_n\hh\hv =\hv_1+\hv_2\,,
\n{4.2}\\
\hv_1(\eta,\theta)&=&-\sum_{n\geq0}
\Big\{3(a_n/2+b_n/2)R^nP_n
+(a_n+b_n/2)\sum_{l=0}^{n-1}(\eta-\cos\theta)R^lP_l\non\\
&+&(a_n/2+b_n)\sum_{l=0}^{n-1}(-1)^{n-l}(\eta+\cos\theta)R^lP_l\Big\}\,,\n{4.3}\\
\hv_2(\eta,\theta)&=&\sum_{n,k\geq1}\frac{nk}{n+k}(a_na_k+a_nb_k+b_nb_k)R^{n+k}
[P_nP_k-P_{n-1}P_{k-1}]\,,\n{4.4}
\ea
where
\be\n{4.5}
R=(\eta^2-\sin^2\theta)^{1/2}
\hh P_n\equiv P_n(\eta\cos\theta/R)\,.
\ee
Here $P_{n}$'s  are the Legendre polynomials of the first kind. These series generally converge if the sources of the distortion fields are located far from the black hole and the fields are considered in the black hole vicinity. Accordingly, as in the Newtonian gravitational theory and electromagnetism, the constant coefficients $a_{n}$'s and $b_{n}$'s, which define the distortion fields, are called {\em interior multipole moments}.

Distortion fields defined by exterior multipole moments correspond to asymptotically flat solutions. On the other hand, according to the uniqueness theorem formulated in \cite{57}, a Schwarzschild-Tangherlini black hole is the only $d-$dimensional, asymptotically flat, static, vacuum black hole which has non-degenerate regular event horizon. This implies that the sources located inside the black hole make its horizon singular. Thus, to have a regular horizon we shall consider non-asymptotically flat solutions distorted by the external sources only, whose distortion fields are defined by the interior multipole moments. Such fields must be regular and smooth at the horizon. The distortion fields $\hu$, $\hw$, and $\hv$ given above satisfy this condition. 

If the sources of the distortion fields are included into the solution, then their energy-momentum tensor satisfies the strong energy condition. The strong energy condition implies that
\be\n{4.11} 
\hu+\hw\leqslant 0\,.
\ee
Then, it follows that, on the ``semi-axes" $\theta=0$ and $\theta=\pi$, at the black hole horizon $\eta=1$,
we have
\be\n{4.12} 
\sum_{n\geq0}(\pm 1)^n(a_{n}+b_{n})\leqslant 0\,,
\ee
where $+1$ corresponds to $\theta=0$ and $-1$ corresponds to $\theta=\pi$.
 
Applying the generating transformation to the metric \eq{4.1} we derive the solution representing distorted charged black hole,
\ba
ds^2&=&-\frac{4p^2(\eta^2-1)}{\Delta^2}e^{2(\hu+\hw)}dt^2+\frac{m\Delta}{2}\left(e^{2(\hv+\hu+\hw)}\frac{d\eta^2}{4(\eta^2-1)}+d\hat{\Omega}^2\right)
\,,\n{4.13}\\
\Delta&=&(1+p)(\eta+1)-(1-p)(\eta-1)e^{2(\hu+\hw)}\,,\n{4.14}\\
d\hat{\Omega}^2&=&\frac{1}{4}\left(e^{2(\hv+\hu+\hw)}d\theta^2
+2(1+\cos\theta)e^{-2\hw}d\chi^2
+2(1-\cos\theta)e^{-2\hu}d\phi^2\right)\,,
\n{4.15}\\
\Phi&=&\frac{\sqrt{3(1-p^2)}}{\Delta}\left(\eta+1-
(\eta-1)e^{2(\hu+\hw)}\right)\,.\n{4.16}
\ea
For $p=1$, this solution represents the distorted five-dimensional Schwarzschild-Tangherlini black hole. If the distortion fields vanish, this solution represents the five-dimensional Reissner-Nordstr\"om solution. Note that, the distorted black hole electric charge $Q$ is independent of the values of the multiple moments and equals to that of the Reissner-Nordstr\"om black hole
\ba
&&ds^2=-\frac{p^2(\eta^2-1)}{(1+p\eta)^2}dt^2+m(1+p\eta)\left(\frac{d\eta^2}{4(\eta^2-1)}+d\Omega^2\right)\,,~~~~
\Phi=\frac{\sqrt{3(1-p^2)}}{1+p\eta},\n{2.14}
\ea
where $d\Omega^2$ is a metric on a unit three-sphere. The parameters $m$ and $q$ are related to the five-dimensional Komar mass of the black hole $M$ and its five-dimensional electric charge $Q$  as follows:
\be\n{2.8}
M=\frac{3\pi}{4}m\hh Q=2\sqrt{3}\,q\,.
\ee
In these coordinates, the event (outer) horizon is at $\eta=1$ and the (inner) Cauchy horizon is at $\eta=-1$, and the space-time singularity is at $\eta=-1/p$. The transformation does not affect the distortion fields $\hu$, $\hw$, and $\hv$. When the distortion fields $\hu$, $\hw$, and $\hv$ vanish, the solution represents a five-dimensional Reissner-Nordstr\"om solution in an empty, asymptotically flat universe. 
One can show that, the space-time curvature invariants diverge in the region where $\Delta=0$, where 
\be\n{4.19}
\eta+\frac{1+p+(1-p)e^{2(\hu+\hw)}}{1+p-(1-p)e^{2(\hu+\hw)}}=0\,.
\ee
Thus, if $\hu+\hw\leqslant 0$, that is if the sources of the distortion fields satisfy the strong energy condition, then the space-time singularities are located behind the inner (Cauchy) horizon. 

In addition, for a regular horizon there should be no conical singularities on the ``semi-axes" $\theta=0$ and $\theta=\pi$, and thus on the horizon. For the metric \eq{4.13}--\eq{4.15} this condition implies (for details see \cite{AS-S})
\be\n{4.6}
\hv+2\hu+\hw|_{\theta=0}=0\,,
\ee 
for the ``semi-axis" $\theta=0$, and 
\be\n{4.7}
\hv+\hu+2\hw|_{\theta=\pi}=0\,,
\ee
for the ``semi-axis" $\theta=\pi$, which can be written in the following form
 \be\n{4.9}
\sum_{n\geq0}(a_{2n}-b_{2n})+3\sum_{n\geq0}(a_{2n+1}+b_{2n+1})=0\,.
\ee
This condition implies the black hole equilibrium condition.
\section{Properties of the distorted black hole solution}

In this section we shall discuss the properties of the distorted black hole solution \eq{4.13}--\eq{4.16}. To begin with, let us introduce the following notations:
\ba
&&u_0=\sum_{n\geq0}a_{2n}\hh u_1=\sum_{n\geq0}a_{2n+1}\hh
w_0=\sum_{n\geq0}b_{2n}\hh w_1=-\sum_{n\geq0}b_{2n+1}\,,\n{5.1}\\
&&u_{\pm}(\theta)=\sum_{n\geq0}(\pm 1)^na_{n}\cos^n\theta-u_0\hh
w_{\pm}(\theta)=\sum_{n\geq0}(\pm 1)^nb_{n}\cos^n\theta-w_0\,.\n{5.2}
\ea
The three-dimensional surface of the outer horizon is defined by $t=const$ and $\eta=1$. The three-dimensional surface of the inner horizon is defined by $t=const$ and $\eta=-1$. One can show that the metrics of the outer and inner horizon surfaces are related to each other by the following transformation, i.e., 
an exchange between the ``semi-axes" and reverse of signs of the multipole moments,
\be\n{5.18}
(\theta,\chi,\phi)\longrightarrow 
(\pi-\theta,\phi,\chi)\hh a_n\longrightarrow-a_n\hh b_n\longrightarrow-b_n\,.
\ee
We shall call the transformation \eq{5.18} the {\em duality transformation} between the outer and inner horizons of the distorted black hole. This transformation is exactly the same as the duality transformation between the horizon and the {\em stretched} singularity surfaces of the distorted five-dimensional Schwarzschild-Tangherlini black hole (see Eqs. (135)--(137), \cite{AS-S}). 
The horizon areas of the distorted black hole solution are 
\be\n{5.19}
\mathcal{A}_\pm=2\pi^2
\sqrt{m^3(1\pm p)^3}\,e^{\mp \frac{3}{2}\gamma}\,.
\ee
One can see that the area product has the same form as that of the Reissner-Nordstr\"om black hole:
\be\n{5.21}
\mathcal{A}_+\mathcal{A}_-=4\pi^4q^3
=\frac{\pi^4}{6\sqrt{3}}Q^3\,.
\ee
We can define the lower and upper limits for the inner and outer horizon areas of a general distorted Reissner-Nordstr\"om black hole,
\be\n{ain2}
\mathcal{A}_{-}<\frac{\pi^2 Q^\frac{3}{2}}{3^{\frac{3}{4}}\sqrt{2}}<\mathcal{A}_{+}\,,
\ee
which can be written in the following form:
\be\n{5.21a}
\mathcal{A}_{-}<\sqrt{\mathcal{A}_{-}
\mathcal{A}_{+}}<\mathcal{A}_{+}\,.
\ee
Thus, the geometric mean of the inner and outer horizon areas of the distorted black hole represents the upper and lower limits of its inner and outer horizon areas, respectively.

It is easy to see that, the values of the electrostatic potential at the black hole horizons does not change under the distortion. One can also check that the Smarr formula 
\be\n{5.22a}
\pm M=\frac{3}{16\pi}\kappa_\pm \mathcal{A}_\pm\pm\frac{\pi}{8}\Phi_\pm Q\,.
\ee
holds for the distorted black hole as well. Here $M$ defines the black hole Komar mass, assuming that the space-time \eq{4.13}--\eq{4.15} can be analytically extended to achieve its asymptotic flatness.

\section{Mechanics and thermodynamics of the distorted black hole}
In this section, we derive mechanical laws of the distorted black hole and present the corresponding laws of thermodynamics. 
The zeroth law says that a black hole surface gravity (and accordingly, its temperature) is constant at the black hole horizon. The surface gravity is defined up to an arbitrary constant which depends on the normalization of the time-like Killing vector. However, the normalization does not affect the zeroth law. The surface gravity at the horizons is
\be\n{5.22}
\kappa_\pm=\frac{2p\,e^{\pm \frac{3}{2}\gamma}}{\sqrt{m(1\pm p)^3}}\,,~~
\gamma=u_0+w_0+\frac{1}{3}(u_1+w_1)\,.
\ee
We see that due to distortion fields, the surface gravity differs from that of the Reissner-Nordstr\"om (undistorted) black hole by the factor $e^{\pm \frac{3}{2}\gamma}$. The zeroth law holds for both horizons of our distorted black hole. The corresponding temperature is defined in terms of the surface gravity as 
\be\n{t1}
T_{\pm}=\frac{\kappa_{\pm}}{2\pi}\,.
\ee
This definition, however, requires a proper normalization of the Killing vector at the spatial infinity. 

Though the temperature $T_{+}$ is associated with the black hole outer horizon is a typical quantity, the ``temperature" $T_{-}$, which is associated with the black hole inner (Cauchy) horizon, is rather a dubious thermodynamic variable. However, taking into account the description of black holes within string theory \cite{Hor}, one can view the inner horizon thermodynamics as the difference of the thermodynamics corresponding to the right- and left-moving excitations of the strings. 

Global first law correspond to the total system of the black hole plus the 
distorting matter. Local first law corresponds to the system of the black hole only. To define a global first law one needs to extend the space-time to achieve its asymptotic flatness. The extension is achieved by requiring that the distortion fields $\hu,\hw$, and $\hv$ vanish  at the asymptotic infinity and by extending the corresponding space-time manifold. In the extended manifold there exists an electrovacuum region in the interior of the black hole and part of the exterior region where the solution \eq{4.13}--\eq{4.16} is valid. Then, there is a region where the  external sources are located. Beyond that region there is asymptotically flat electrovacuum region. Having this extension one can normalize the timelike Killing vector $\BM{\xi}_{(t)}$ at the spatial infinity as $\BM{\xi}_{(t)}^2=-1$. As it is done, one  naturally finds that
the komar mass M of undistorted black hole correspond to the mass of 
black hole not including the mass of external matter. We derive the global first law of the black hole mechanics,
\be\n{t2}
\pm\delta M=\frac{\kappa_\pm}{8\pi}\delta \mathcal{A}_\pm
\pm\frac{\pi}{8}\Phi_\pm\delta Q+M_\pm^{\text{loc}}\delta\gamma\,,
\ee
where the local black hole mass,
\be\n{t3}
M_\pm^{\text{loc}}=\pm\frac{3\pi}{4}mp\,,
\ee
does not depend on the distortion fields. 
From the global first law of the black hole mechanics, by using the definition of temperature \eq{t1} and the black hole entropy,
\be\n{t4}
S_\pm=\frac{\mathcal{A}_\pm}{4}\,,
\ee
we derive the global first law of the black hole thermodynamics,
\be\n{t5}
\pm\delta M=T_\pm\delta S_\pm
\pm\frac{\pi}{8}\Phi_\pm\delta Q+M_\pm^{\text{loc}}\delta\gamma\,.
\ee       
Here the term $M_\pm^{\text{loc}}\delta\gamma$ is interpreted as the work done on the black hole by the variation of the external potential $\gamma$ due to the distorting matter. If the distortion is adiabatic, $\delta S_\pm=0$, i.e., such that neither matter nor gravitational waves cross the black hole horizons, and in addition, the black hole charge $Q$ does not change, then the work $M_\pm^{\text{loc}}\delta\gamma$ results in the change of the black hole mass $\delta M$. 

The local first law does not include the distorting matter into consideration of the black hole mechanics. The observers who live near the black hole consider the black hole as an isolated, undistorted object. Thus, assuming that there is no other matter present and the space-time is asymptotically flat, they define its surface gravity $\tilde{\kappa}_+$, the outer horizon area $\mathcal{\tilde A}_+$, electrostatic potential $\tilde{\Phi}_+$, electric charge $\tilde{Q}$, and the black hole Komar mass $\tilde{M}$. Thus, these observers construct the local first law of the black hole mechanics as that of the Reissner-Nordstr\"om (undistorted) black hole, 
\be\n{t7}
\pm\delta \tilde{M}=\frac{\tilde{\kappa}_\pm}{8\pi}\delta \mathcal{\tilde{A}}_\pm
\pm\frac{\pi}{8}\Phi_\pm\delta Q\,.
\ee
With the definitions of temperature \eq{t1} and entropy \eq{t4} the local first law of black hole thermodynamics reads
\be\n{t8}
\pm\delta \tilde{M}=\tilde{T}_\pm\delta \tilde{S}_\pm
\pm\frac{\pi}{8}\Phi_\pm\delta Q\,.
\ee 
The measurements of the observers define the black hole area as that which is exactly equal to the black hole area when the presence of the distortion fields is taken into account, i.e.,
\be\n{t9}
\mathcal{\tilde{A}}_\pm=2\pi^2
\sqrt{\tilde{m}^3(1\pm \tilde{p})^3}=\mathcal{A}_{\pm}=2\pi^2
\sqrt{m^3(1\pm p)^3}\,e^{\mp \frac{3}{2}\gamma}\,,
\ee
where 
\be\n{t10}
\tilde{M}=\frac{3\pi}{4}\tilde{m}\hh \tilde{p}=\frac{1}{\tilde{m}}\sqrt{\tilde{m}^2-q^2}\,.
\ee
The following relations provide us with the correspondence between the local and the global forms of the first law:
\ba
\tilde{M}&=&\frac{M}{2}\left[(1+p)e^{-\gamma}+(1-p)e^{\gamma}\right]\,,\n{t11}\\
\tilde{p}&=&\frac{(1+p)e^{-\gamma}-(1-p)e^{\gamma}}{(1+p)e^{-\gamma}+(1-p)e^{\gamma}}\,,\\
\tilde{\kappa}_{\pm}&=&\frac{2\tilde{p}}{\sqrt{\tilde{m}(1\pm\tilde{p})^{3}}}=\frac{\kappa_{\pm}}{2p}\left[(1+p)e^{-\gamma}-(1-p)e^{\gamma}\right]\,,\n{t13}\\
\tilde{\Phi}_{\pm}&=&\frac{\sqrt{3(1-\tilde{p}^2)}}{1\pm \tilde{p}}=\Phi_{\pm}e^{\gamma}\,,\n{t14}
\ea
\section{Conclusion}
 We have constructed a distorted, five-dimensional Reissner-Nordstr\"om black hole solution. We have studied the mechanic and thermodynamic of the solution. In addition, this black hole posses some other interesting properties, which we haven't mentioned here. We observed, that  the space-time singularities are located behind the black hole's inner (Cauchy) horizon, provided that the sources of the distortion satisfy the strong energy condition. The inner (Cauchy) horizon remains regular if the distortion fields are finite and smooth at the outer horizon. There exists a certain duality transformation between the inner and the outer horizon surfaces which links surface gravity, electrostatic potential, and space-time curvature invariants calculated at the black hole horizons. The product of the inner and outer horizon areas depends only on the black hole's electric charge and the geometric mean of the areas is the upper (lower) limit for the inner (outer) horizon area. The electromagnetic field invariant calculated at the horizons is proportional to the squared surface gravity of the horizons. The horizon areas, electrostatic potential, and surface gravity satisfy the Smarr formula. We formulated the zeroth and the first laws of mechanics and thermodynamics of the distorted black hole and found a correspondence between the global and local forms of the first law. To illustrate the effect of distortion we can consider the dipole-monopole and quadrupole-quadrupole distortion fields. This would provide us with the result that the relative change in the Kretschamnn scalar due to the distortion is greater at the outer horizon than at the inner one. By calculating the maximal proper time of free fall from the outer to the inner horizons one can show that the distortion can noticeably change the black hole interior. The change depends on type and strength of distortion fields. In particular, due to the types of distortion fields considered the black hole horizons can either come arbitrarily close to or move far from each other.
 \acknowledgments

S. A. gratefully acknowledges the Deutsche Forschungsgemeinschaft (DFG) for financial support within the framework of the DFG Research Training group 1620 Models of gravity.

\end{document}